\documentclass{ecai} 

\usepackage{graphicx}
\usepackage{latexsym}

\usepackage{graphicx}
 \usepackage{amsmath}
\usepackage{latexsym}
\usepackage{amssymb} 
\usepackage{tikz}
\usetikzlibrary{shapes.geometric, arrows, patterns}
\usepackage{subcaption}
\usepackage{caption}
\usepackage{wrapfig}
\usepackage{tabularx}
\usepackage{ragged2e}
\usepackage{color, colortbl}
\definecolor{Gray}{gray}{0.9}

\tikzstyle{stegblock} = [rectangle, rounded corners, minimum width=1.5cm, minimum height=1cm,text centered, draw=black, fill=green!30]
\tikzstyle{decision} = [ellipse, minimum width=1.5cm, minimum height=1cm, text centered, draw=black]
\tikzstyle{arrow} = [thick,->,>=stealth]

\usepackage{xcolor}

\newcommand{\cover}[0]{C}
\newcommand{\secret}[0]{S}

\newcommand{\container}[0]{C'}
\newcommand{\revealsecret}[0]{S'}
\newcommand{\revealsanisecret}[0]{\hat{S}}
\newcommand{\sanitized}[0]{\hat{C}}
\newcommand{\sanitizedNoise}[0]{\hat{C}_{\mathit{noise}}}
\newcommand{\sanitizedSUDS}[0]{\hat{C}_{\mathit{SUDS}}}
\newcommand{\Reveal}[0]{\mathcal{R}}
\newcommand{\Reveallsb}[0]{\mathcal{R}_{\mathit{lsb}}}
\newcommand{\Revealddh}[0]{\mathcal{R}_{\mathit{ddh}}}
\newcommand{\Revealudh}[0]{\mathcal{R}_{\mathit{udh}}}
\newcommand{\Hide}[0]{\mathcal{H}}
\newcommand{\features}[0]{n}
\newcommand{\latentvar}[0]{\mathbf{z}}
\newcommand{\diff}[0]{\mathit{diff}}
\newcommand{\dataset}[0]{D}
\newcommand{\inputim}[0]{x}
\newcommand{\logvar}[0]{\boldsymbol{\log\sigma^2}}

\newcommand{\kl}[0]{\mathit{KL}}
\newcommand{\muv}[0]{\boldsymbol{\mu}}
\newcommand{\enc}[0]{\mathit{Enc}}
\newcommand{\dec}[0]{\mathit{Dec}}

\newcommand{\figref}[1]{Figure~\ref{#1}}
\newcommand{\tabref}[1]{Table~\ref{#1}}
\newcommand{\secref}[1]{Section~\ref{#1}}

\begin{document}

\begin{frontmatter}

\title{SUDS: Sanitizing Universal and Dependent \\ Steganography}

\author[A]{\fnms{Preston K.}~\snm{Robinette}\thanks{Corresponding Author. Email: preston.k.robinette@vanderbilt.edu}}
\author[A]{\fnms{Hanchen D.}~\snm{Wang}}
\author[A]{\fnms{Nishan}~\snm{Shehadeh}}
\author[A]{\fnms{Daniel}~\snm{Moyer}}
\author[A]{\fnms{Taylor T.}~\snm{Johnson}}

\address[A]{Vanderbilt University}

\begin{abstract}
Steganography, or hiding messages in plain sight, is a form of information hiding that is most commonly used for covert communication. As modern steganographic mediums include images, text, audio, and video, this communication method is being increasingly used by bad actors to propagate malware, exfiltrate data, and discreetly communicate. Current protection mechanisms rely upon steganalysis, or the detection of steganography, but these approaches are dependent upon prior knowledge, such as steganographic signatures from publicly available tools and statistical knowledge about known hiding methods. These dependencies render steganalysis useless against new or unique hiding methods, which are becoming increasingly common with the application of deep learning models. To mitigate the shortcomings of steganalysis, this work focuses on a deep learning sanitization technique called SUDS that is not reliant upon knowledge of steganographic hiding techniques and is able to sanitize universal and dependent steganography. SUDS is tested using least significant bit method (LSB), dependent deep hiding (DDH), and universal deep hiding (UDH). We demonstrate the capabilities and limitations of SUDS by answering five research questions, including baseline comparisons and an ablation study. Additionally, we apply SUDS to a real-world scenario, where it is able to increase the resistance of a poisoned classifier against attacks by 1375\%.
\end{abstract}

\end{frontmatter}

\section{Introduction}
Steganography is the art of hiding information in plain sight in order to discreetly communicate \cite{Johnson_steg, kessler_steg}. Deriving from the Greek words ``steganos'' (covered) and ``grafia'' (writing), steganography literally means covered writing, and it is prevalent throughout much of history. Whereas cryptography uses encryption to make a message incomprehensible to the naked eye, steganography hides the traces of the communication entirely. By human nature, an encrypted message attracts attention and incites scrutiny. People are attracted to the allure of breaking a cypher and are often successful in breaking encryption keys given enough time and computational resources. Steganography, however, has the benefit of escaping the scrutiny of unassuming eyes, offering a discreet method of communication which can be used to hide both encrypted and unencrypted information, making it a potentially dangerous attack vector for bad actors. 

Steganography has been used across several creative mediums, including books, knitting, and wax tablets. Modern approaches most commonly rely on digital media such as images, audio, text, and videos. As digital media is easily distributed and widely spread, the potential effects of steganography have grown exponentially, making it important to be able to protect against this type of communication if used by bad actors. While steganography can be used for applications like watermarking proprietary information \cite{luo2020distortion} and light field messaging \cite{Wengrowski_lfm}, attackers can leverage this communication technique to propagate malware \cite{suarez2015stegomalware}, exfiltrate victim data \cite{kaspersky_exfil}, and communicate.

In an effort to limit the adverse effects of steganography, recent research has focused on steganalysis, or the detection of steganography. If, for instance, an advertisement containing a malware payload is detected on a web browser, this image can be removed from the site, protecting clients from interacting with these malicious images. Current steganalysis techniques rely on statistical image tests and signatures associated with known steganography techniques \cite{Montasari2023, bhme2010advanced, pevny2009steganalysis, gul2011new, fridrich2012rich, shi2013textural}. Additionally, deep learning steganographic techniques utilize datasets of images created with publicly available steganography tools. Existing detection methods, therefore, rely on the assumption that a hiding signature or information about a steganography technique is already known. What happens, then, as new, not publically available hiding methods begin to surface, making current detection techniques irrelevant? While an increasingly difficult problem to address, detection should not be abandoned. Instead, it would be more advantageous to use detection in conjunction with other protective mechanisms, providing a more robust protection system.

As such, this work focuses on a sanitization framework for image steganography. The sanitization of an image eliminates any hidden information within the image while keeping the quality of the actual image the same. Whereas detection is used to identify steganographic images, sanitization mitigates the use of steganography entirely. Drawing upon research related to denoising images, this work uses a variational autoencoder framework to create a deep learning model that \underline{\textbf{s}}anitizes \underline{\textbf{u}}niversal and \underline{\textbf{d}}ependent \underline{\textbf{s}}teganography, called SUDS. To the best of the authors' knowledge, SUDS is the first sanitization framework which can mitigate the effects of traditional, dependent deep, and universal deep hiding. The contributions of this work, therefore, are the following:
\begin{enumerate}
    \item \textbf{Implementation of a Robust Sanitizer.} We construct and train a framework capable of sanitizing traditional, dependent deep, and universal deep hiding steganography techniques, called SUDS.
    \item \textbf{Demonstration of Sanitizer Capabilities} We show the benefit of this novel framework by evaluating SUDS on five capabilities: ability to sanitize, comparison to noise, flexibility of the latent dimension (ablation study), ability to detect, and scalability.
    \item \textbf{Case Study Application.} We demonstrate a use case of SUDS whereby SUDS is able to protect against data poisoning, increasing classifier resistance against attacks by 1375\%. 
\end{enumerate}

\section{Preliminaries}
\label{Sec:prelims}
 While steganography can be used across any medium, this work refers to image steganography, where an image is an $(c, h, w)$ matrix, where $c$ is the number of color channels, $h$ is the height of the image, and $w$ is the width. An RGB image is then represented by an $(3, h, w)$ matrix and a grayscale image by an $(1, h, w)$ matrix.

\subsection{Steganography Nomenclature}
\begin{table}
\begin{center}
{\caption{Steganography Notation}\label{tab:steg_terms}}
\begin{tabular}{|>{\Centering}p{.2\columnwidth}>{\Centering}p{.15\columnwidth}>{\RaggedRight}p{.45\columnwidth}|}
\hline
\rule{0pt}{10pt}
\textbf{Term} &  \textbf{Symbol} & \textbf{Definition} \\
\hline
\rule{0pt}{10pt}
Cover &$\cover$ & The image used to hide (or cover up) a secret. A cover is combined with a secret to create a container.\\
\rule{0pt}{10pt}
    Secret & $\secret$ & The image to be hidden. \\
\rule{0pt}{10pt}
    Container & $\container$ & A cover image that contains a secret. A container should look identical to the cover used to conceal the secret. \\
\rule{0pt}{10pt}
    Revealed Secret & $\revealsecret$ & The secret revealed from the container. The revealed secret should be identical to the actual secret.\\
\rule{0pt}{10pt}
    Sanitized & $\sanitized$ & A sanitized image. \\
\rule{0pt}{10pt}
    Revealed Sanitized Secret & $\revealsanisecret$ & The secret revealed from a sanitized image. \\
\rule{0pt}{10pt}
    Hide & $\Hide$ & A method to hide a secret within a cover.\\
\rule{0pt}{10pt}
    Reveal & $\Reveal$ & A method to reveal the hidden secret within a container.\\
\rule{0pt}{10pt}
    Difference & $\diff$ & The difference calculation between images. This work uses the $\mathcal{L}_2$ norm, also known as the mean squared error. \\
\hline
\end{tabular}
\end{center}
\end{table}
Steganography usually consists of four components: a cover $\cover$, a secret $\secret$, a container $\container$, and a revealed secret $\revealsecret$ \cite{zhang2020udh, baluja2017hiding} as described in \tabref{tab:steg_terms}. A hiding method $\Hide$ takes as input a cover and a secret and produces a container such that the visible difference between the original cover and the produced container is minimal. Written more formally, $\Hide(\cover, \secret) = \container$ s.t. $\diff(\container, \cover)$ is minimized, where $\diff$ in this work is the $\mathcal{L}_2$ norm. A reveal function $\Reveal$ then takes as input the produced container and outputs the revealed secret $\revealsecret$ such that the visible difference between the original secret and the revealed secret is minimal, or $\Reveal(\container) = \revealsecret$ s.t. $\diff(\revealsecret, \secret)$ is minimized. An example of this steganographic process is depicted on the \textit{Pre-Sanitization} side of \figref{fig:suds}. In this work, we utilize $\sanitized$ to represent a sanitized image such that the image is sanitized with either noise or SUDS, i.e., $\sanitized \in \{\sanitizedNoise, \sanitizedSUDS\}$, and $\revealsanisecret$ to represent a secret revealed with $\Hide$ from a sanitized image, i.e., $\Hide(\sanitized) = \revealsanisecret$.

\subsection{Steganography Hiding Methods}
\begin{figure}
\centering
\begin{tikzpicture}
\node (coveruse) [decision] {Dependent on Cover?};
\node (dnnuse) [decision, below of=coveruse, xshift=-2cm, yshift=-0.7cm] {Use of Neural Network?};
\node (udh) [stegblock, below of=coveruse, xshift=2cm, yshift=-0.7cm] {UDH};
\node (ddh) [stegblock, below of=dnnuse, xshift=-2cm, yshift=-0.7cm] {DDH};
\node (lsb) [stegblock, below of=dnnuse, xshift=2cm, yshift=-0.7cm] {LSB};
\draw [arrow] (coveruse) -- node[anchor=east] {yes} (dnnuse);
\draw [arrow] (coveruse) -- node[anchor=west] {no} (udh);
\draw [arrow] (dnnuse) -- node[anchor=east] {yes} (ddh);
\draw [arrow] (dnnuse) -- node[anchor=west] {no} (lsb);
\end{tikzpicture}
\caption{\centering Dichotomy of steganography used in this work.}
\label{fig:steg_flowchart}
\end{figure}
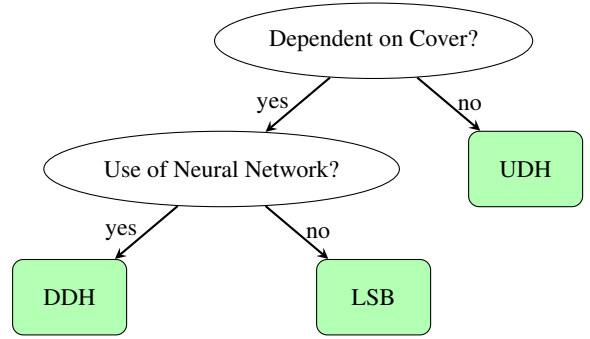

Existing methods for steganography currently fall into three main categories: traditional, dependent deep, and universal deep. We use one from each category to highlight the robustness of the proposed sanitization approach: traditional $\to$ Least Significant Bit Method (LSB) \cite{kurak_CSAC}, dependent deep $\to$ Dependent Deep Hiding (DDH) \cite{volkhonskiy2020steganographic, yang2019embedding, tang2017automatic, tang2020automatic, wu2020gan, zhu2018hidden, wang2018sstegan,baluja2017hiding}, and universal deep $\to$ Universal Deep Hiding (UDH) \cite{zhang2020udh}. We utilize our own implementation of LSB\footnote{\textbf{SUDS  Code:} \url{https://github.com/pkrobinette/suds-ecai-2023}} and a CNN-based implementation of DDH and UDH from the same code base\footnote{\textbf{DDH/UDH: }\url{https://github.com/ChaoningZhang/Universal-Deep-Hiding}}, which was chosen due to its dual functionality. The hyperparameters used during UDH and DDH training are shown in \tabref{tab:hparams}. The main distinctions between these methods are highlighted in \figref{fig:steg_flowchart}, and we leave detailed descriptions of these methods to the behest and curiosity of the reader.

\subsection{Sanitization}
Sanitization is a method of removing an intended secret from a container image. It renders the message of the container incomprehensible. Whereas detection is used to identify steganographic images, sanitization mitigates the use of steganography entirely.

\textbf{Noise Baseline} A way to disrupt steganographic messages is to add noise to images. This approach not only degrades the visual rendering of the image, but it is also not a robust method for sanitization. To create a baseline of sanitization, we apply \textit{Gaussian} noise to containers in this work. An image sanitized by noise is referred to as $\sanitizedNoise$, i.e., $\sanitizedNoise = \mathit{clip}(\inputim + \mathcal{N}(\mu, \sigma), \mathit{min}=0, \mathit{max}=1)$, where $\mu=0$, $\sigma=0.02$, and the $\mathit{clip}$ function keeps altered pixels in a meaningful range. The $\mu$ and $\sigma$ values were chosen to mimic the maximum change of using LSB, which is approximately $6\%$ for the four least significant bits $(\frac{15}{255}*100 = 6\%)$. 

\begin{figure*}[htbp!]
    \centering
    \includegraphics{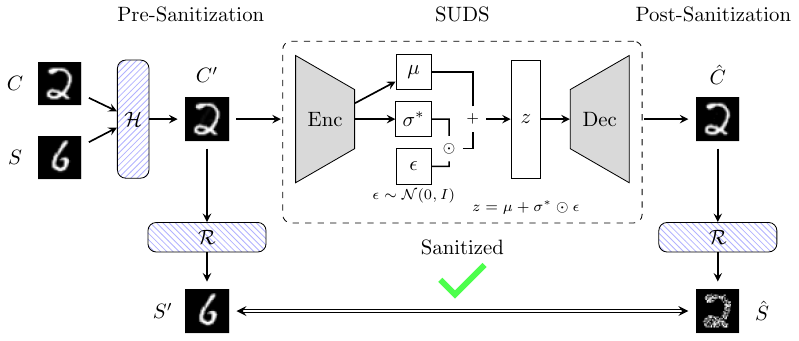}
    \caption{\centering SUDS overview. After being trained, SUDS takes as input a cover or a container image $\inputim\in\{\cover, \container\}$ created with any type of steganographic technique. Prior to being sanitized, the secret is recoverable, as demonstrated in the bottom-left of the figure $\revealsecret$. After sanitization with SUDS, however, the reconstructed image $\sanitized$ still looks the same as the input $\container$, but a secret is not recoverable as indicated by the bottom-right of the figure $\revealsanisecret$. This image, therefore, is successfully sanitized.}
    \label{fig:suds}
\end{figure*}

\textbf{Variational Autoencoder} This work utilizes a variational autoencoder (VAE) as the framework for the sanitization model SUDS, as shown in \figref{fig:suds}. VAEs are a type of representation learning that consist of an encoder and a decoder. 
\label{sec:vae_info}
\begin{itemize}
    \item \textbf{Encoder:} The encoder $\enc$ takes as input an input image $\inputim$ from a dataset $\dataset$. The input image $\inputim \in \mathbb{R}^{\mathit{chw}}$ is mapped to two different vectors of feature size $\features$, representing the mean $\muv \in \mathbb{R}^{\mathit{n}}$ and logarithm of the variance $\logvar \in \mathbb{R}^{\mathit{n}}$ of $\features$ distributions, i.e., $\enc(\inputim) =  \muv, \logvar$. A latent variable is then sampled from $\muv$ and $\logvar$ to derive a latent vector $\latentvar \in \mathbb{R}^n$ as shown in Equation \eqref{z_sample}, where $\odot$ is the element-wise product and $\epsilon\sim\mathcal{N}(0,I)$.  The inclusion of $\odot\epsilon$ in \eqref{z_sample} is known as the \textit{reparametrization trick}, and it allows gradients to be computed and backpropagated through the network. 
    
        \begin{equation}
        \latentvar = \mathit{f}(\muv, \logvar) = \muv + e^{\frac{1}{2}\logvar}\odot\epsilon
        \label{z_sample}
        \end{equation}

    An encoder, therefore, maps an input image to a latent representation: $\enc(\inputim) =  \muv, \logvar \implies \latentvar = \mathit{f}(\muv, \logvar)$ s.t. $\mathbb{R}^{chw} \to \mathbb{R}^{n}$. The depiction of this process in \figref{fig:suds} shows only $\sigma^*$ for simplicity. While $e^{\frac{1}{2}\logvar} = \sigma$, the output of $\enc$ in our implementation is represented as the logarithm of the variance to simplify the computation of the Kullback-Leibler divergence term discussed in \secref{sec:suds_algo}. 
    
    \item \textbf{Decoder:} Once mapped to a latent variable $\latentvar$, the decoder $\dec$ can then create a reconstruction of the original image $\hat{\inputim} \in \mathbb{R}^{chw}$ from this condensed representation, i.e., $\dec(\latentvar) = \hat{\inputim}$ s.t. $\mathbb{R}^{n} \to \mathbb{R}^{chw}$. The goal of the decoder is to reconstruct an image such that $\diff(\inputim, \hat{\inputim})$ is minimal where $\diff$ is the mean-squared error or the $\mathcal{L}_2$ norm.
\end{itemize}

\noindent In this work, input images are either a cover or container $\inputim \in \{\cover, \container\}$, and the reconstructed image $\hat{x} = \sanitizedSUDS$, where $\sanitizedSUDS$ is an image sanitized by SUDS. An image ``sanitized'' by SUDS is then an image that has been encoded and decoded by the SUDS framework. 

\subsection{Image Metrics}
In this work, we utilize mean-squared error (MSE) and peak-signal-to-noise ratio (PSNR) image metrics described in Equations \eqref{eq:mse} and \eqref{eq:psnr} respectively, where $A$ and $B$ are the compared images of size (c, h, w) and $\mathit{MAX}$ is the maximum possible pixel value (for a given bit depth).

\begin{equation}
\mathit{MSE}(A,B)= \frac{1}{chw} \sum_{i=1}^{c} \sum_{j=1}^{h} \sum_{k=1}^{w} (A_{i,j,k} - B_{i,j,k})^2
\label{eq:mse}
\end{equation}

\begin{equation}
\mathit{PSNR}(A, B) = 10 \log_{10} \left(\frac{\mathit{MAX}^2}{\mathit{MSE}(A,B)}\right)
\label{eq:psnr}
\end{equation}


\section{Sanitizing Universal and Dependent Steganography}
\label{sec:approach}
We aim to mitigate the use of image steganography to disseminate embedded secrets while preserving the integrity of the cover image through sanitization. As detection methods rely on signatures from known steganographic techniques, sanitization attempts to render the secret irretrievable, regardless of the steganographic technique used to hide the secret.

\subsection{Sanitization via Variational Autoencoder}

In order to address steganography sanitization, we utilize a variational autoencoder (VAE) to encode images into learned representations, which can then be decoded into a clean counterpart. We propose this solution as motivated by two properties related to VAEs: robustness (encoder) and expressiveness (decoder). 

\textbf{Robustness.} In regard to the robustness of the encoder, a steganographic image can be considered an error value added to the cover, $\container = \cover + \epsilon$. If the VAE is robust, then any container within some bound of the cover will be mapped to the same distribution of the cover, i.e., $\enc(\cover + \epsilon)  = \enc(\cover)  \: \text{s.t.} \: |\epsilon| < \gamma$, where $\enc$ is an encoder, $\epsilon$ is an error value, and $\gamma$ is an arbitrary error bound. In other words, covers and containers should be mapped to the same latent representation if the model is robust. 

\textbf{Expressiveness.} Whereas robustness is related to the encoder,  expressiveness is related to the decoder. By encoding a potentially steganographic image into a condensed representation, we create an information bottleneck. This property helps to bound possible decoded images, which in turn decreases the potential to decode an image from the representation which contains an embedded secret. There are more possible outcomes from a network that relies on 784 pieces of information (the original image) than from a network that uses only 128 pieces of information (the encoded latent space). The latent space representation, therefore, limits the expressiveness of the decoder, which aids in the sanitization process.

\subsection{SUDS Framework Overview}
\label{sec:suds_overview}
Building upon these concepts, we propose SUDS, a VAE framework which is able to \textbf{\underline{s}}anitize \textbf{\underline{u}}niversal and \textbf{\underline{d}}ependent \textbf{\underline{s}}teganography. As shown in \figref{fig:suds}, SUDS takes as an input a container or a cover and produces a sanitized version of the input. Prior to sanitization, the secret of the container $\container$ is recoverable, as indicated by the revealed secret $\revealsecret$ in the bottom-left of the image. After sanitization, the intended secret is no longer recoverable, and the image quality is minimally different from that of the original cover used to create the container. The attempted recovery of the secret $\hat{\secret}$ shown in the bottom-right of the image is not the intended secret $\secret$, rendering the attempted communication useless and the image successfully sanitized.

\subsection{SUDS Algorithm Details}
\label{sec:suds_algo}
During training, SUDS takes as input covers $\cover$ from a dataset $\dataset$. Each input image is mapped to a latent variable of size $n$, as described in \secref{sec:vae_info}. For instance, if $\features = 128$, the latent variable will consist of 128 features sampled from Equation \eqref{z_sample}. To create a smooth and continuous latent space, we encourage the learned distribution to resemble a normal distribution by adding a regularizing term to the loss function, the Kullback-Leibler (KL) divergence, as shown in \eqref{loss_regularization}. Here, $q_{\phi}(z|x)$ is the conditional probability distribution of the latent variable $\latentvar$ given a cover $\cover$ and $\mathcal{N}(0, I)$ is the induced normal distribution. The KL divergence is a measure of how different one probability distribution is compared to another.

\begin{equation}
        L_{reg} = \kl(q_{\phi}(\latentvar|\cover)|| \mathcal{N}(0, I))
        \label{loss_regularization}
\end{equation}

Once mapped to a latent variable $\latentvar$, the encoded cover image is then decoded to a reconstruction of the original image, referred to as $\sanitizedSUDS$. The reconstruction loss associated with the decoder is shown in Equation \eqref{loss_reconstruction}, where $\mathbb{E}_{q_{\phi}(z|\cover)}$ is expectation given the approximate posterior distribution of the latent variable $\latentvar$ given the input data $\cover$, $\sanitized$ is the reconstructed output of the decoder, and $\diff$ is the $\mathcal{L}_2$ norm.

\begin{equation}
        L_r = \mathbb{E}_{q_{\phi}(\latentvar|\cover)}{[\diff(\cover, \sanitizedSUDS)]}
        \label{loss_reconstruction}
\end{equation}

The encoder and decoder are trained in tandem by combining their regularization and reconstruction loss functions, as shown in Equation \eqref{loss_total}. Training occurs for a specified number of epochs, where one epoch consists of all batched training data. It is important to note that SUDS is not trained on steganographic images, only clean images. 

\begin{equation}
        \mathcal{L} = L_r + L_{reg}
        \label{loss_total}
\end{equation}

\begin{figure*}
    \centering
    \includegraphics[width=\linewidth]{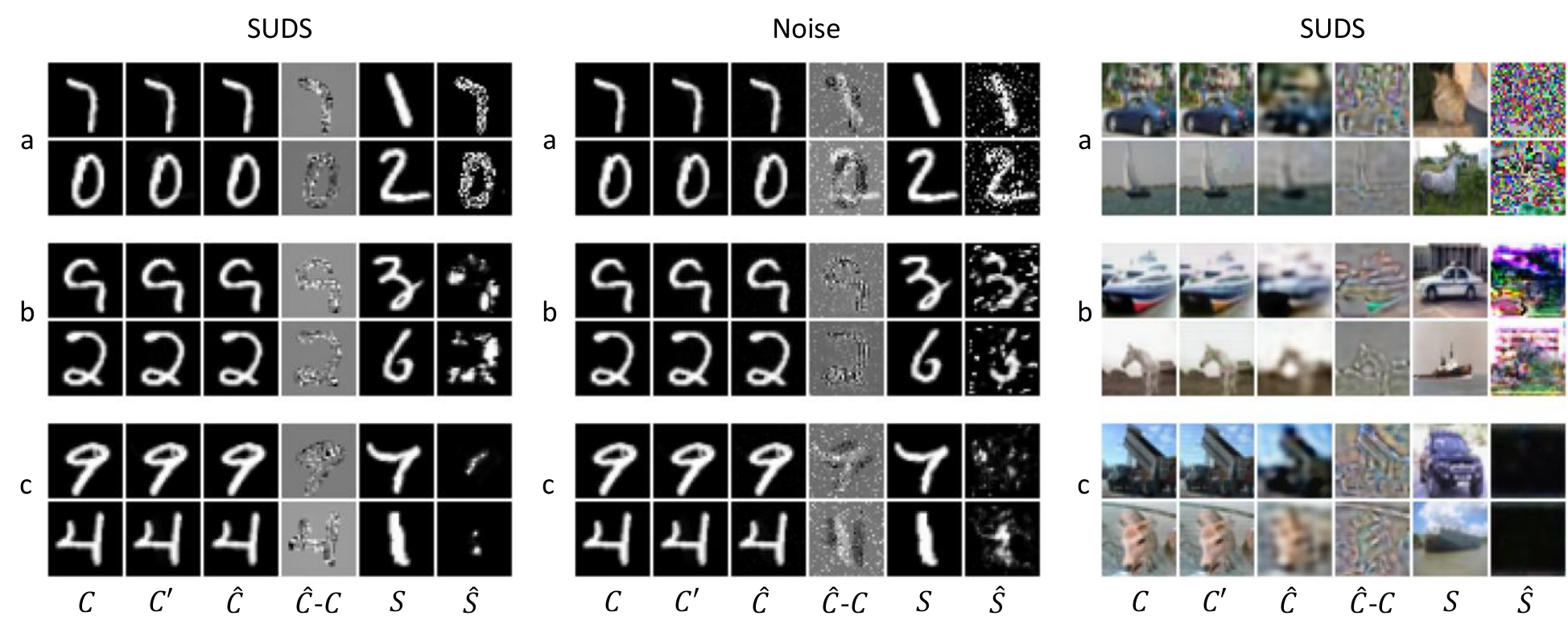}
    \caption{\centering Image results for SUDS and noise sanitization on containers hidden with \textbf{a)} LSB, \textbf{b)} DDH, \textbf{c)} UDH.  A cover $\cover$ is combined with a secret $\secret$ to create a container $\container$. The container image is then sanitized with SUDS $\sanitizedSUDS$ or noise $\sanitizedNoise$ as indicated by the group title. A sanitizer is successful if $\secret$ is not discernible from $\revealsanisecret$. SUDS, therefore, is successful at sanitizing, while noise is not.}
    \label{fig:results}
\end{figure*}

\section{Research Questions and Metrics}

To evaluate SUDS, we seek to answer five questions:

\begin{enumerate}
    \item \textbf{RQ1: Ability to sanitize steganography $\to$} \textit{Is SUDS able to remove the presence of potential secrets while preserving the integrity of the cover?} To answer this question, we first train SUDS using the MNIST dataset as described in \secref{sec:suds_algo} and with the hyperparameters shown in \tabref{tab:hparams}. The MNIST training dataset consists of 60000 images $(1, h, w)$ of handwritten digits from numbers 0-9. After training, we then use 10000 images from the MNIST test dataset to create containers using each of the hiding methods, i.e. LSB, DDH, UDH. These containers are then sanitized with SUDS and passed to each reveal function. If the intended secret is discernible with any of the reveal functions, the sanitizer is not effective at cleaning images. Additionally, we evaluate SUDS using peak signal-to-noise ratio (PSNR) and mean squared error (MSE) metrics, which are commonly used image metrics. These metrics are used to assess a sanitizer's effect on the input ($x \text{ compared to } \sanitized$ s.t. $x \in \{\cover, \container\}$) as well as its ability to sanitize hidden secrets ($\secret \text{ compared to } \revealsanisecret$). \textbf{RQ1} is positive if the revealed secret after sanitization $\revealsanisecret$ is not the intended secret $\secret$, i.e.,  $\revealsanisecret \not\approx \secret$ where $\revealsanisecret = \Reveal(\sanitizedSUDS)$.
    
    \item \textbf{RQ2: Comparison to noise baseline $\to$} \textit{How does SUDS compare to noise baseline techniques?} We repeat the testing process of \textbf{RQ1} but replace SUDS with Gaussian noise to sanitize the containers. We then analyze the resulting noise MSE and PSNR values against those produced by SUDS. \textbf{RQ2} is positive if SUDS is able to outperform noise in MSE and PSNR values.

    \item \textbf{RQ3: Flexibility of the Latent Dimension $\to$} \textit{Does latent dimension size affect performance?} We train SUDS on different latent dimension sizes $n$ to see if and when SUDS is no longer able to sanitize a steganographic image. Using the same training framework described in \secref{sec:suds_algo}, this case study tests $n \in \{2, 4, 8, 16, 32, 64, 128\}$ and evaluates each SUDS model using MSE and PSNR image quality metrics. \textbf{RQ3} is positive if a majority ($\geq 4$) of the trained latent dimension sizes of SUDS are able to sanitize secrets, as indicated by MSE and PSNR values.

    \item \textbf{RQ4: Ability to detect steganography $\to$} \textit{In addition to sanitization, can SUDS be used to detect steganography?} We use 10000 MNIST images of varying image types, i.e., cover images, LSB containers, DDH containers, and UDH containers. These images are then mapped to a latent variable $z$ by using SUDS trained on $n=8$ features. The containers for each hiding method are created using the same cover and secret across each method. To evaluate detection capabilities, we calculate the mean and standard deviation for each image type and digit. For instance, if Label = 4, the resulting data is calculated across all covers with the label 4 for each image type. \textbf{RQ4} is positive if the distributions across a single digit are distinguishable.

    \item \textbf{RQ5: Scalability $\to$} \textit{Does SUDS scale to RGB images?} We use the CIFAR-10 dataset, which consists of 60000  RGB images $(3, 32, 32$) from 10 different classes of various animals and transportation vehicles. To train SUDS on CIFAR-10, we use the same training setup detailed in \secref{sec:suds_algo}. \textbf{ RQ5} is positive if the revealed secret after sanitization $\revealsanisecret$ is not the intended secret $\secret$, i.e.,  $\revealsanisecret \not\approx \secret$ where $\revealsanisecret = \Reveal(\sanitizedSUDS)$.
    
\end{enumerate}

\begin{table}
\begin{center}
{\caption{\centering Model Hyperparameter Values}\label{tab:hparams}}
    \centering
    \renewcommand{\arraystretch}{1.05}
    \begin{tabular}{|c|cc|}
    \hline
    \textbf{Model} & \textbf{Hyperparameter} & \textbf{Value} \\
    \hline
    & epochs & 100 \\
    & $\features$ & 128 \\
    SUDS & batch size & 128 \\
    & optimizer & Adam \\
    & learning rate & 0.0001 \\
    \hline
    & image size & 32  \\
    & batch size & 44 \\
    & channel of cover & 1  \\
    DDH/UDH & channel of secret & 1  \\
    & norm & ``batch''  \\
    & loss & ``l2'' \\
    & beta & 0.75   \\
    \hline
\end{tabular}
\end{center}
\end{table}

\section{Experimental Results}
The results of the SUDS sanitizer training and evaluation are described in more detail below. These experiments were conducted on a macOS Monterey 12.5.1 with a 2.3 GHz 8-Core Intel Core i9 processor with 16 GB 2667 MHz DDR4 of memory.
\subsection{Sanitization Performance via SUDS}
\label{sec:sani_perf}
Using the evaluation approach described in \textbf{RQ1}, the sanitization framework (SUDS) is able to effectively clean each of the container images so that the revealed secret is not obtainable through any of the reveal functions while keeping the original cover intact. As shown by the left-hand group in \figref{fig:results},  $\sanitizedSUDS$ and $\cover$ have limited variation, and $\secret$ is not retrieved by $\Revealddh(\sanitizedSUDS)$, $\Revealudh(\sanitizedSUDS)$, or $\Reveallsb(\sanitizedSUDS)$ as shown by $\revealsanisecret$.

The PSNR and MSE metrics shown at the top of \tabref{tab:sani_results} further validate SUDS. The MSE and PSNR values of the \textit{Clean} image provide a baseline for the steganographic images. The MSE and PSNR values of the \textbf{$\sanitized$} column compare an input image $x \in \{\cover, \container\}$ to a sanitized image $\sanitized\in\{\sanitizedNoise, \sanitizedSUDS\}$, i.e., $\sanitized_{\mathit{MSE}} = \mathit{MSE}(\inputim, \sanitized)$ and $\sanitized_{\mathit{PSNR}} = \mathit{PSNR}(\inputim, \sanitized)$. A small MSE value indicates minimal changes between the input and the output image, and a high PSNR value indicates that the signal outweighs the noise in the image. SUDS is, therefore, able to reproduce input images with a high quality regardless of whether the image is a container or cover. The \textbf{$\revealsecret$} column evaluates the reveal functions of each hide method before the container has been sanitized. The MSE and PSNR values of this column compare to the revealed secret prior to sanitization to the intended secret, i.e., $\revealsecret_{\mathit{MSE}} = \mathit{MSE}(\secret, \revealsecret)$ and $\revealsecret_{\mathit{PSNR}} = \mathit{PSNR}(\secret, \revealsecret)$. As shown by the small MSE and high PSNR values, each of the reveal functions is effectively able to derive the intended secret such that $\secret \approx \revealsecret$. In the \textbf{$\revealsanisecret$} column, however, the intended secret is not discernible from a sanitized image. The MSE and PSNR values for this column compare the intended secret to the secret revealed after sanitization, i.e., $\revealsanisecret_{\mathit{MSE}} = \mathit{MSE}(\secret, \revealsanisecret)$ and $\revealsanisecret_{\mathit{PSNR}} = \mathit{PSNR}(\secret, \revealsanisecret)$. The high MSE values and low PSNR values confirm that the intended secret is not obtainable through the reveal functions after sanitization. Thus, SUDS is effectively able to sanitize steganographic images, and \textbf{RQ1 is positive}.

\label{sec:suds_results}
\begin{table}
\begin{center}
{\caption{\centering Image metrics for SUDS and noise sanitization methods (\secref{sec:sani_perf}). Bolded $\revealsanisecret$ values mean successful sanitization.}\label{tab:sani_results}}
\centering
\begin{tabular}{|ccc|ccc|}
    \hline
    \rule{0pt}{10pt}
    & & & $\sanitized$ & $\revealsecret$ & $\revealsanisecret$\\
\hline
        & Clean & MSE  & 0.28& - & - \\ 
        &       & PSNR & 53.77& - & - \\ 
\cline{2-6}
        & LSB & MSE & 0.45& 0.09 & \textbf{62.72}\\ 
        &     & PSNR & 51.72& 58.46 & 30.17\\ 
\cline{2-6}
 SUDS  & DDH & MSE & 0.33& 0.49 & \textbf{73.45}\\ 
        &     & PSNR & 53.03& 51.73 & 29.47\\ 
\cline{2-6}
        & UDH & MSE & 0.38& 0.63 & \textbf{88.54}\\ 
        &     & PSNR & 52.36& 50.23 & 28.66\\ 
\hline
        & Clean & MSE  & 0.09& - & - \\ 
        &       & PSNR & 58.82& - & - \\ 
\cline{2-6}
        & LSB & MSE & 0.09& 0.09 & 30.04\\ 
        &     & PSNR & 58.49& 58.46 & 33.37\\ 
\cline{2-6}
 Gaussian  & DDH & MSE & 0.09& 0.49 & 38.1\\ 
        &     & PSNR & 58.5& 51.76 & 32.33\\ 
\cline{2-6}
        & UDH & MSE & 0.12& 0.63 & \textbf{57.58}\\ 
        &     & PSNR & 57.35& 50.26 & 30.53\\ 
\hline
\end{tabular}
\end{center}
\end{table}

\subsection{Performance Comparison to Noise Baseline}
\label{sec:baseline_results}
In addition to SUDS, we also test a Gaussian noise baseline, as described in \textbf{RQ2}. \tabref{tab:sani_results} shows the image quality metric performance for the noise baseline sanitization technique. As indicated by the bolded value in the \textbf{$\revealsanisecret$} column, Gaussian noise is only effective at sanitizing secrets hidden with UDH, and is therefore not a robust solution. This is further validated by the middle group in \figref{fig:results}, which shows the image results of using noise to sanitize. In the \textbf{$\revealsanisecret$} column, we can clearly see the secrets for both LSB and DDH. As a result, SUDS outperforms Gaussian noise for sanitization, and \textbf{RQ2 is positive}.

Another benefit to SUDS compared to noise lies in its ability to get closer to the original cover. When noise is added as a sanitization technique, the image quality is degraded and the image is pushed further from the actual rendering of the original image. With SUDS, however, as it is trained to reconstruct cover images, the resulting sanitized image gets closer to the original value. In this way, SUDS works to ``improve'' an image's quality, as shown by the $\sanitized - \cover$ for each respective method in \figref{fig:results}.

\subsection{Effect of Latent Dimension Size}
\begin{figure}[htbp!]
        \centering
        \includegraphics[width=\columnwidth]{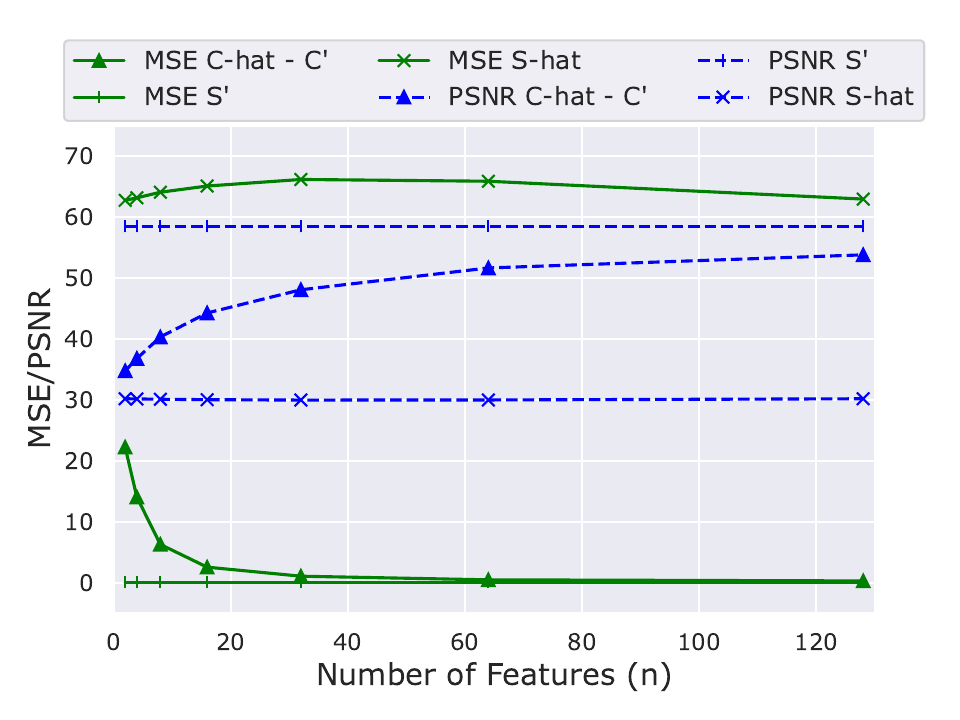}
        \caption{\centering MSE and PSNR values for SUDS trained on varying feature sizes ($n$). SUDS is able to sanitize at each feature size as indicated by the top green-x line, but its reconstruction ability deteriorates as feature size decreases as shown by $n = 4$ and $n = 2$ on the green-triangle line.}
        \label{fig:models_plot}
\end{figure}

SUDS also works with a variety of feature space sizes. For \textbf{RQ3}, we train SUDS on different feature sizes $n$ to see if and when SUDS is no longer able to sanitize a steganographic image. As shown in \figref{fig:models_plot}, SUDS retains its ability to sanitize a secret, as indicated by the high green-x line for each feature size. The sanitization performance peaks between $n=64$ and $n=32$. While the sanitization performance is latent-size agnostic, SUDS's ability to reconstruct images starts to deteriorate as $n$ decreases, as shown by the high MSE values of the green-triangle line and the low PSNR values of the blue-triangle line for $n=4$ and $n=2$. This result makes sense; as the number of features decreases, the amount of information passed to the decoder also decreases, making it difficult to accurately reconstruct the original image. These results show that \textbf{RQ3 is positive}.


\subsection{Evaluation of Detection Capabilities}
\label{sec:latent_space}
\begin{figure}[htbp!]
     \centering
         \centering
         \includegraphics[width=\columnwidth]{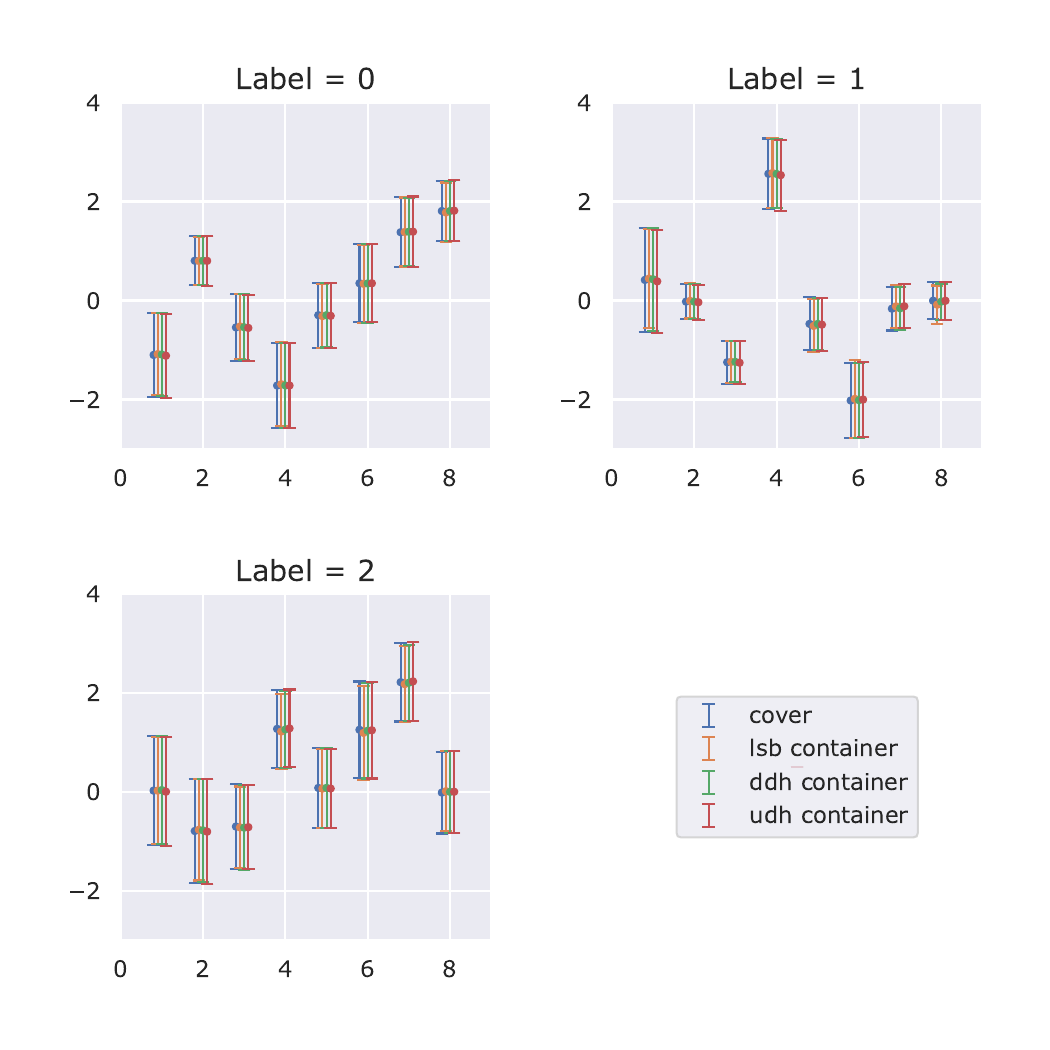}
         \caption{\centering The mean $\mu$ and standard deviation $\sigma$ values for images of a particular label ($\text{label} \in \{0, 1, 2, ..., 9\}$) mapped to a latent space representation ($z$) from SUDS. The mean is represented by the center dot of each line, and the standard deviation is represented by the error bars of each line. From these results, the input images are not distinguishable from the latent space.}
         \label{fig:latent_space}
\end{figure}

In addition to analyzing the impact of feature size on SUDS performance, we also evaluate whether SUDS can be used as a detection mechanism by analyzing where images get mapped to in the feature space. From the results shown in \figref{fig:latent_space}, each image, regardless of the presence of a secret, is mapped to relatively the same distribution per feature. As there is no distinction between covers and containers as well as hiding methods in the feature space, SUDS cannot be used for a detection mechanism based on mean and standard deviation metrics. Thus, \textbf{RQ4 is negative}.

\subsection{Scalability to RGB Images}

SUDS is able to sanitize RGB images as well as grayscale images, as indicated by the right-hand side of \figref{fig:results}. Once again, the intended secret $S$ is not discernible from $\revealsanisecret$. Image reconstruction capabilities, however, decline with RGB images, as shown by $\sanitized$. We believe that this performance can be increased by additional hyperparameter tuning, architecture tuning, and increased training epochs. From these results, \textbf{RQ5 is positive}.

\subsection{Summary}
In summary, \textbf{RQ1} (ability to sanitize), \textbf{RQ2} (comparison to noise), \textbf{RQ3} (flexibility), and \textbf{RQ5} (scalability) are all positive while \textbf{RQ4} (detection) is negative.

\section{Example Application: Data Poisoning}
To demonstrate a potential use case, we apply SUDS to a case study involving data poisoning. Data poisoning is an adversarial machine learning attack where a model is perturbed during training. This can be caused by methods such as data injection and most commonly---data manipulation. In data manipulation, an attacker has access to the data used for training, and they can alter, add, or remove training data as well as data labels to create backdoors in a model. Common forms of data manipulation include pattern injection, single-pixel modifications, and steganography, which are visibly hard to detect---especially in the case of steganography. Although there are many methods to manipulate data for data poisoning, this case study highlights the use of SUDS to mitigate a data poisoning attack that utilizes steganographic perturbations.


\textbf{Overview.} Data poisoning usually occurs in three phases: preparation, training, and attack. Each of these phases is described in more detail below.

\begin{enumerate}
    \item \textbf{Preparation}: The data perturbation method is created, and the data to be used during training is modified using this technique. 
    \item \textbf{Training}: The model is trained using the manipulated training data.
    \item \textbf{Attack}: The poisoned model is deployed. The attackers can then craft an adversarial example to alter the intended behavior of the poisoned model.
\end{enumerate}

For this case study, the MNIST training dataset of 60000 images is prepared using the deep dependent hiding (DDH) technique to perturb images. Using DDH, 40\% of the training data are changed to a container image, where the secret is randomly chosen with replacement from the input set. The label of this container image is then modified to the label of the randomly chosen secret hidden within it. A classifier is then trained using this manipulated data to classify the digit shown in the picture. If the data poisoning scheme is successful, an attacker should be able to direct the prediction of the trained network by selecting the desired classification as the secret of the container image input. 

This data poisoning technique is evaluated using 10000 test images, where 50\% have been changed to container images and the remaining 50\% are kept clean. The data poisoning technique is then tested again by incorporating SUDS between the client and the trained network. In this test, the data distribution is identical to the previous test, but all test data is sanitized prior to being classified with the trained network.

\textbf{Result.} \tabref{tab:poison_results} shows the classification accuracy during testing of both clean and container images with and without the use of SUDS. The accuracy associated with the container images indicates the classification of the intended secret to what was predicted by the classifier---was the attacker successful in manipulating the classifier to the intended secret? For instance, if an attacker hid a secret image of label 6 in a cover image with a label of 4, the model would be accurate if it predicted the secret label of 6. By using SUDS between the client and the trained model, the resistance of the classifier to an attack increases by $1375\% = \frac{(100-0.56) - (100-93.26)}{100-93.26}*100$, while the performance for clean images only drops by 1\%. SUDS, therefore, is successfully able to protect the classification system from bad actors.

\begin{table}
\renewcommand{\arraystretch}{1.15}
\begin{center}
{\caption{\centering Classification accuracy of clean and poisoned images, both with using SUDS to protect against poisoned images and without using SUDS (no SUDS).}\label{tab:poison_results}}
\begin{tabular}{|>{\Centering}p{.35\columnwidth}|>{\Centering}p{.25\columnwidth}>{\Centering}p{.25\columnwidth}|}
    \hline
     & \multicolumn{2}{c|}{\textbf{Accuracy}} \\
    \cline{2-3}
         Image Type & no SUDS & SUDS  \\
         \hline
         clean (5000) & 98.03 \% & 97.18\% \\
         containers (5000) & 93.26\% & 0.56 \% \\
    \hline
\end{tabular}
\end{center}
\end{table}

\section{Related Works}
While SUDS is the first sanitization framework that can mitigate the effects of traditional, dependent deep, and universal deep hiding to the best of the authors' knowledge, there are similar works which use VAEs for protection. In \cite{zuppelli2021sanitization}, the authors show the promise of using a variational autoencoder (VAE) to mitigate the transfer of PowerShell scripts hidden directly in images with Invoke-PSImage \cite{zuppelli2021sanitization}, a publicly available hiding tool that uses least significant bit hiding\footnote{\textbf{Invoke-PSImage:} https://github.com/peewpw/Invoke-PSImage}. In this previous work, the sanitizer is evaluated on whether the resulting steganographic image is detected by StegExpose, an open-source detection tool for the least significant bit hiding methods. StegExpose, though, is not a robust detection tool. Additionally, even small perturbations in scripts hidden in images can cause a script to become non-executable. As images are more robust to perturbations, evaluating sanitization with image secrets provides better evaluation criteria for this approach. Additionally, while a sanitizer may be able to sanitize steganography resulting from one hiding technique, this does not mean that it will be effective in sanitizing steganography with other hiding techniques as each method and signature is unique. 

A VAE has also been used to detect and repair adversarial perturbations, which are modifications to inputs to cause misclassification. In \cite{meng2017magnet}, the authors implement Magnet, which uses outlier detection in the latent space to determine if an image is adversarial. As adversarial techniques add noise, resulting images are mapped to different parts of the latent space. During the repair phase, the outlier latent representations are perturbed in the direction up the gradient to get closer to the distribution of the actual training data, working to remove the added noise of the adversarial image. While steganography is similar to an adversarial image, the ``noise'' of a steganographic image, or the message, is bounded, i.e., the message must be retrievable and make sense to the recipient. The bounded nature of steganography results in images that are mapped to the same distribution in the latent space (see \secref{sec:latent_space}) and, therefore, are not applicable to the Magnet approach.

\section{Conclusions}
In this work, we demonstrate a sanitization model SUDS, which is able to sanitize steganographic images from untrained steganographic techniques (traditional, dependent deep, and universal deep) with more reliability than baseline sanitization methods. Whereas SUDS sanitizes secrets for all hiding techniques used in this work, the baseline noise sanitization technique only works 33\% of the time against these hiding techniques. In addition to performing more reliably compared to noise, SUDS also has the added benefit of producing a sanitized image that is closer to that of the original cover used to make a container, while noise degrades an image in order to render the hidden secret incomprehensible. This allows SUDS to be applied to many interesting case studies, such as protection against data poisoning attacks, increasing resistance of a poisoned classifier against attacks by 1375\%. In the future, we would like to extend SUDS to different steganographic mediums, such as videos, binaries, and time series data, and test this approach against other representation learning models such as a generative adversarial network or diffusion models. As such, this research lays the groundwork for future exploration in comprehensive steganography protection systems.

\ack This paper was supported in part by a fellowship award under contract FA9550-21-F-0003 through the National Defense Science and Engineering Graduate (NDSEG) Fellowship Program, sponsored by the Air Force Research Laboratory (AFRL), the Office of Naval Research (ONR), and the Army Research Office (ARO). The material presented in this paper is based upon work supported by the National Science Foundation (NSF) through grant numbers 2220426 and 2220401, the Defense Advanced Research Projects Agency (DARPA) under contract number FA8750-23-C-0518, and the Air Force Office of Scientific Research (AFOSR) under contract number FA9550-22-1-0019 and FA9550-23-1-0135. Any opinions, findings, and conclusions or recommendations expressed in this paper are those of the authors and do not necessarily reflect the views of AFOSR, DARPA, or NSF.

\bibstyle{ecai}
\bibliography{references}

\end{document}